\documentclass[showpacs,preprintnumbers, twocolumn]{revtex4-2}

\usepackage[colorlinks]{hyperref}% add hypertext capabilities

\usepackage[centertags]{amsmath}
\usepackage{amsfonts}
\usepackage{amssymb}
\usepackage{amsthm}
\usepackage{newlfont}
\usepackage{graphicx}
\usepackage{lipsum}
\usepackage{comment}

\usepackage{tikz}
\usetikzlibrary{calc}
\usepackage{tikz-3dplot}
\usetikzlibrary{patterns}
\usetikzlibrary{shadows}

\begin{document}

\newcommand{\pp}[1]{\phantom{#1}}
\newcommand{\be}{\begin{eqnarray}}
\newcommand{\ee}{\end{eqnarray}}
\newcommand{\M}[1]{{\color{magenta}{#1}}}
\newcommand{\Y}[1]{{\color{yellow}{#1}}}
\newcommand{\C}[1]{{\color{cyan}{#1}}}
\newcommand{\G}[1]{{\color{green}{#1}}}
\newcommand{\R}[1]{{\color{red}{#1}}}
\newcommand{\B}[1]{{\color{blue}{#1}}}

\title{Relativity of spacetime ontology: When correlations in space become correlata in time}
\author{Marek Czachor}

 \email[Electronic address: ]{marek.czachor@pg.edu.pl}

\author{Marcin Nowakowski}
 \email[Electronic address: ]{marcin.nowakowski@pg.edu.pl}

\affiliation{Faculty of Applied Physics and Mathematics, Gdańsk University of Technology, 80-952 Gdańsk, Poland}
\affiliation{National Quantum Information Center, 80‑309 Gdańsk, Poland}


\begin{abstract}

 Challenging Mermin's perspective that ``correlations have physical reality; that which they correlate does not'' we argue that correlations and correlata are not fundamentally distinct. These are dual concepts depending on the tensor product decomposition defining subsystems. Since the same quantum states may be either entangled or separable, but with respect to alternative tensor product structures, a spatial correlation in one context can become a temporal correlatum in another, and vice versa. In consequence, 2-qubit states invariant under $V\otimes V$ can be either entangled or unentangled, in conflict with the well known uniqueness theorem about the singlet state, a fact with possible implications for the quantum measurement theory.

\end{abstract}

\maketitle

\section*{Introduction}

The roles of the observer and the nature of correlations are foundational and yet often enigmatic both in quantum mechanics and relativity theory. Traditional perspectives, such as those posited by Mermin, suggest that while correlations possess physical reality, the elements they correlate do not. This paper challenges this dichotomy, proposing instead that correlations and correlata are not inherently distinct but are dual aspects influenced by observers defining subsystems and their correlations, i.e. defining elements of reality.

Our inquiry, grounded in standard quantum mechanics, implies that spatial correlations in one context may transform into temporal correlata in another, indicating that space-time itself may not be a fundamental level of reality.

To begin with, consider an invertible map $V$ that describes a change of basis (or a reference frame) in a single-qubit system. 
A two-qubit state,
$
|\psi\rangle
=
\sum_{\alpha\beta}\psi_{\alpha\beta}|\alpha\rangle\otimes |\beta\rangle,
$
is invariant under $V\otimes V$,   
$
(V\otimes V)|\psi\rangle\sim|\psi\rangle,
$
if and only if 
$
|\psi\rangle \sim |0\rangle\otimes |1\rangle-|1\rangle\otimes |0\rangle.
$ 
This uniqueness theorem is behind various quantum and relativistic constructions, just to mention spinor reformulation of Einstein-Maxwell equations \cite{PR}, or proofs of security of a two-qubit entangled-state quantum cryptography \cite{BBM92}.

In particular, if $V\in \textrm{SU}(2)$, $|0_V\rangle=V|0\rangle$, $|1_V\rangle=V|1\rangle$, then
the Bell-basis singlet state satisfies
\be
|\Psi_-\rangle &=& \frac{1}{\sqrt{2}}\big(|0\rangle\otimes |1\rangle-|1\rangle\otimes |0\rangle\big)
\label{Intro 4}
\\
&=& \frac{1}{\sqrt{2}}\big(|0_V\rangle\otimes |1_V\rangle-|1_V\rangle\otimes |0_V\rangle\big),
\label{Intro 5}
\ee
and, up to a phase factor, is the only state with such a property.

Security of a two-qubit quantum cryptography is based on the following consequence of the uniqueness theorem: Alice and Bob can share singlet-state correlations if and only if they are isolated from the rest of the universe. But then, if we replace Alice and Bob by spins of electrons in a helium atom, how is it that one observes singlet state correlations in atomic physics? The two electrons must be isolated from the rest of the universe, so apparently cannot be subject to spectroscopic measurements.

The question brings us to another controversial and general property of quantum mechanics ---  the quantum measurement problem \cite{WZ,BLM}. Namely, a system is described by a state vector $|\psi_s\rangle$ if and only if the overall  state of the universe is a tensor product state, $|\psi_s\rangle\otimes |\psi_r\rangle$, with $|\psi_r\rangle$ representing the rest of the universe. But then the system must be isolated from anything else and is therefore essentially unobservable, so how is it that any quantum mechanical prediction based on the Schr\"odinger equation is testable at all?

The goal of our paper is to show that the above controversies may be rooted in a fundamental non-uniqueness of tensor product structures \cite{Zanardi,Kus,Lloyd,Caban,Bartlett,Thirring,Ahmad}. A two-qubit system is naturally equipped with six coexisting tensor products $\otimes_{abc}$, indexed by the six permutations of $a,b,c=1,2,3$. The tensor products then satisfy, in particular,
\be
|\Psi_-\rangle &=& \frac{1}{\sqrt{2}}\big(|0\rangle\otimes_{123} |1\rangle-|1\rangle\otimes_{123} |0\rangle\big)
\label{Intro 6}\\
&=&
\frac{1}{\sqrt{2}}\big(|0_V\rangle\otimes_{123} |1_V\rangle-|1_V\rangle\otimes_{123} |0_V\rangle\big)
\label{Intro 7}\\
&=&
\frac{1}{\sqrt{2}}|1\rangle\otimes_{321} \big(|1\rangle-|0\rangle\big)
\label{Intro 8}\\
&=&
\frac{1}{\sqrt{2}}|1_V\rangle\otimes_{321} \big(|1_V\rangle-|0_V\rangle\big),
\label{Intro 9}
\ee
for any $V\in \textrm{SU}(2)$. The last equality in (\ref{Intro 8})--(\ref{Intro 9}) is in apparent conflict with the discussed uniqueness theorem. However, we will see that (\ref{Intro 8}) equals (\ref{Intro 9}) for any $V$ in virtue of the theorem, not in spite of it.

Our text may be also treated as belated polemic with David Mermin's ``What is quantum mechanics trying to tell us?'' \cite{Mermin1}. In Mermin's own words, his paper can be reduced to a single statement: {\it Correlations have physical reality; that which they correlate does not\/}. ``That which they correlate'' is termed by Mermin the correlata (plural form of correlatum). So, equivalently: {\it Correlations have physical reality; correlata do not\/}.

Our analysis implies that {\it correlations are a form of correlata, but with respect to a different splitting into subsystems\/}. Accordingly, there is no fundamental difference between correlata and correlations. If correlata do not possess physical reality, the same applies to correlations. 

We argue as follows. Correlations describe relations between subsystems. Subsystems are defined in terms of tensor products. In a two-qubit case, subsystem qubits are defined by the four projectors: $P_{\alpha}\otimes \mathbb{I}$ and $\mathbb{I}\otimes P_{\alpha}$, $\alpha=0,1$. However, even in this simple scenario there exist at least three  physically inequivalent tensor products. Depending on our choice of $\otimes$, physical interpretations of $P_{\alpha}$ change. In some cases an eigenvalue of $P_{\alpha}$ describes a correlatum, but in the other cases it describes a correlation. 

In particular, we formally show that correlations in space can become correlata in time, and the other way around. A space-time structure occurs here implicitly as information encoded in pre- or post-selection. It could be further formalized if one replaced qubits by 2-spinors \cite{MC1997,MC2003,MC2008}, and included entanglement in time \cite{Nowakowski2017,Nowakowski2018}, interference in time \cite{MC2019a}, or  swapping space for time \cite{MC2019b} --- the generalizations beyond the scope of the present paper. 

We purposefully restrict our discussion to the simplest case of just two qubits. Many of the usual subtleties related to tensor-product ambiguities are here absent. For example, those related to different factorizations of the dimension of the Hilbert space \cite{Zanardi,Lloyd,Ahmad} disappear trivially because $4=2\times 2$ is effectively the only decomposition (as oposed to, say, $8=2\times 2\times 2=4\times 2$). It is thus easier to extract the ontological and ``causal'' element we are interested in. 
Last but not least, pairs of 2-spinors are enough to define a space-time structure.

\section*{Results}

\subsection*{Correlations and only correlations?}\label{Correlations}
 Mermin posits \cite{Mermin1} that physical reality of a system is entirely encapsulated within (a) the correlations among its subsystems and (b) its correlations with other systems when perceived collectively as components of a more extensive system. He denotes these as the internal and external correlations of the system, respectively. A system is regarded as completely isolated if it lacks external correlations. Furthermore, the wave function of a physical system or, more generally, its
quantum state (pure or mixed) is a concise encapsulation of its internal
correlations \cite{Mermin1}.

One of the foundational presumptions in the so-called Ithaka interpretation of quantum mechanics is the assertion that ``correlations possess physical reality, while that which they correlate does not.''

This poses a conundrum: if that which they correlate does not form a part of physical reality, then how can it be considered a component of reality or how correlations exist between non-existent parts of reality? Furthermore, we delve deeper to illustrate that correlations can be identified among other correlations, while correlations between correlations and correlata automatically accompany correlations occurring between the correlata themselves. Accordingly, following Mermin's line of reasoning, one could claim that the subjects of these correlations (i.e. other correlations) do not exist as elements of physical reality, which contradicts the supposition about the existence of correlations. 

We show that \textit{what can be formally classified as an element of reality — a correlation within one tensor product — can be regarded as a correlatum within another tensor product}. Correlata are integral components of correlations and are always defined in terms of these correlations. As such, correlations can be perceived as a dual concept to correlata. 
In this context, when considering spatial and temporal quantum correlations, we conclude that spatial correlations become temporal correlata,  hinting at a foundational role of correlations in the genesis of space-time.

\subsection*{Tensor products or subsystems? --- quantum chicken-or-egg dilemma}

One usually begins with well defined subystems, whose Hilbert spaces are spanned by two orthonormal vectors. The composite system is spanned by their tensor products,
\be
|00\rangle &=& |0\rangle \otimes |0\rangle ,\label{1?}\\
|01\rangle &=& |0\rangle \otimes |1\rangle ,\\
|10\rangle &=& |1\rangle \otimes |0\rangle ,\\
|11\rangle &=& |1\rangle \otimes |1\rangle.
\ee
The bases at both sides of $\otimes$ can be different.
In order to define yes-no observables corresponding to the two qubits, one first defines the four projectors,
\be
P_{00} &=& |00\rangle\langle 00|=|0\rangle\langle 0|\otimes |0\rangle\langle 0|=P_0\otimes P_0,\\
P_{01} &=& |01\rangle\langle 01|=|0\rangle\langle 0|\otimes |1\rangle\langle 1|=P_0\otimes P_1,\\
P_{10} &=& |10\rangle\langle 10|=|1\rangle\langle 1|\otimes |0\rangle\langle 0|=P_1\otimes P_0,\\
P_{11} &=& |11\rangle\langle 11|=|1\rangle\langle 1|\otimes |1\rangle\langle 1|=P_1\otimes P_1.\label{8?}
\ee
Their averages represent joint probabilities $p_{\alpha\beta}$ of finding correlata $\alpha$ and $\beta$,
\be
p_{\alpha\beta}=\langle\psi|P_{\alpha\beta}|\psi\rangle.
\ee
The probabilities of $\alpha$ or $\beta$ alone,
\be
p^{(1)}_{\alpha} &=&\sum_{\beta} p_{\alpha\beta},\label{10!}\\
p^{(2)}_{\beta} &=&\sum_{\alpha} p_{\alpha\beta},\label{11!}
\ee
are obtained as averages of the following projectors:
\be
P^{(1)}_{\alpha} &=&\sum_{\beta} P_{\alpha\beta}=\sum_{\beta} P_{\alpha}\otimes P_{\beta}=P_{\alpha}\otimes \mathbb{I},\\
P^{(2)}_{\beta} &=&\sum_{\alpha} P_{\alpha\beta}=\sum_{\alpha} P_{\alpha}\otimes P_{\beta}=\mathbb{I}\otimes P_{\beta}.
\ee
Formulas
\be
P_0\otimes \mathbb{I} &=& P_{00}+P_{01},\label{14!}\\
P_1\otimes \mathbb{I} &=&  P_{10}+P_{11},\\
\mathbb{I}\otimes P_0 &=& P_{00}+P_{10},\\
\mathbb{I}\otimes P_{\beta} &=& P_{01}+P_{11},\label{17!}
\ee
can be elegantly represented as sums in rows (the left bit) or columns (the right bit) of the matrix
\be
\left(
\begin{array}{cc}
P_{00} & P_{01}\\
P_{10} & P_{11}
\end{array}
\right).
\label{matrix}
\ee
Matrices with rows or columns interchanged do not define different subsystems (they generate local {\tt NOT} operations, $0\leftrightarrow 1$, in the subsystems). 

Let us note that the right-hand sides of (\ref{10!})--(\ref{11!}) define probabilities of the results 0 and 1 in the detectors of Alice and Bob on the basis of joint detections at both detectors. So, our starting point is the composite system, but constructed with the help of our prior knowledge of $\otimes$ and the form of the subsystems.

However, and this is essential for the argument we advocate in the paper, formulas  (\ref{14!})--(\ref{17!}) provide an operational definition of the tensor product itself. It is sufficient to know the form of $P_{\alpha\beta}$ to define single-qubit observables. The intermediate stage of constructing the composite system can be skipped. An experimental physicist may even forget about the very concept of a tensor product.

Yet, when we consider the issue of completeness of quantum mechanics we have to work at a more general level. We should begin with the composite system treated as some four dimensional Hilbert space, and then define subsystems {\it and\/} tensor products. We use here the plural form because one will not arrive at a unique tensor product, but at six different products, which can be reduced to three if we treat permutation of Alice and Bob as the same decomposition into subsystems. 

Each of these products has a clear physical interpretation. What is important, a correlatum for one decomposition, becomes a correlation in the other decomposition.

Ignoring the multitude of coexisting tensor products we make quantum mechanics incomplete.

\subsection*{Six tensor product structures of two qubits}
\begin{figure}
\includegraphics[width=4 cm]{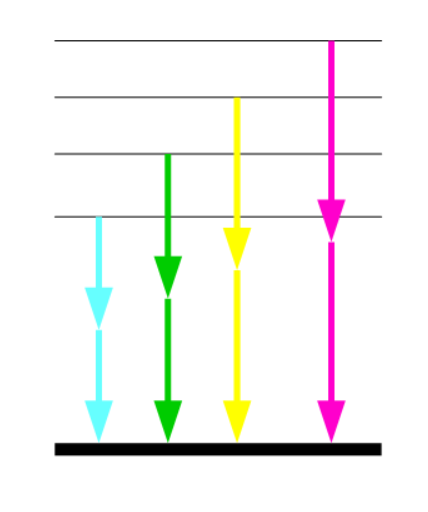}
\caption{A single unstable system in superposition of four orthogonal states decays into a pair of particles. Products of the decay are directed toward the labs of Alice and Bob.}
\label{Fig1}
\end{figure}
\begin{figure}
\includegraphics[width=5 cm]{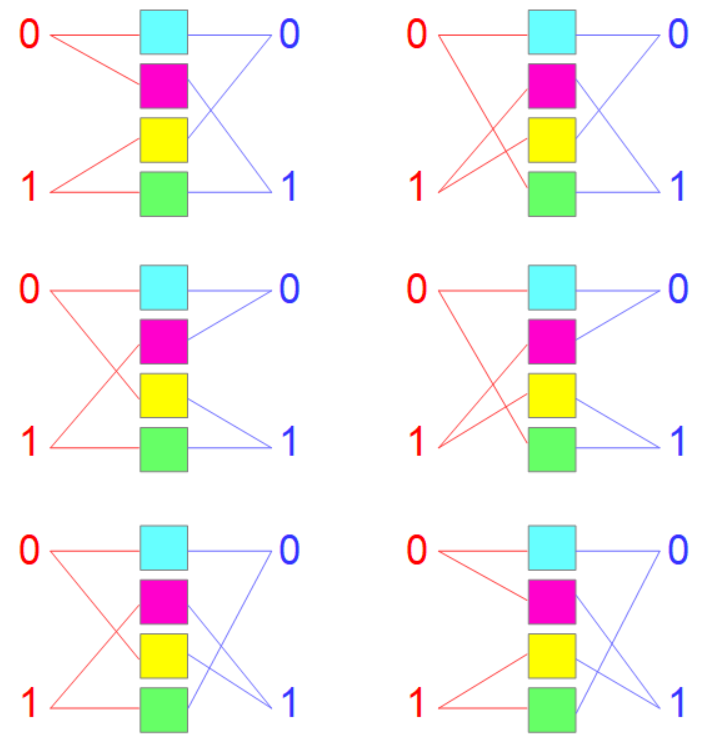}
\caption{Products of the decay can be decomposed into six inequivalent subsystems by erasing information about the color of an emitted particle, and connecting pairs of output channels with single detectors. Bits of Alice are here red, and those of Bob are blue.  Each of the combinations defines a different tensor product structure, effectively labeling the colors by numbers in binary notation. The left column: $(C,M,Y,G)=(00,01,10,11)$ (top), $(00,10,01,11)$ (middle), $(00,11,01,10)$ (bottom). The right column: $(00,11,10,01)$ (top), $(00,10,11,01)$ (middle), $(00,01,11,10)$ (bottom). The remaining permutations of colors lead to local interchanges of bits, $\R{0}\leftrightarrow \R{1}$ or $\B{0}\leftrightarrow \B{1}$.}
\label{Fig6}
\end{figure}

Consider a two-particle decay process involving  a four-dimensional Hilbert space of outgoing states. The four output channels are  denoted in Fig.~\ref{Fig1} by the four CYGM colors: cyan ($C$),  yellow ($Y$), green ($G$), and magenta ($M$). The output consists of two-particle states, so we would like to define two 1-particle subsystems. 

With each of the colors we associate a projector, $P_C=|C\rangle\langle C|$, etc. Having four projectors we can write matrices of the form (\ref{matrix}). Taking into account 
that matrices with rows or columns interchanged do not define different subsystems, we are left with six possibilities:
\begin{align}
&\left(
\begin{array}{cc}
P_C & P_M\\
P_Y & P_G
\end{array}
\right)
&\left(
\begin{array}{cc}
P_C & P_G\\
P_Y & P_M
\end{array}
\right)\label{4,}
\\
&\left(
\begin{array}{cc}
P_C & P_Y\\
P_M & P_G
\end{array}
\right)
&\left(
\begin{array}{cc}
P_C & P_G\\
P_M & P_Y
\end{array}
\right)
\\&\left(
\begin{array}{cc}
P_C & P_Y\\
P_G & P_M
\end{array}
\right)
&\left(
\begin{array}{cc}
P_C & P_M\\
P_G & P_Y
\end{array}
\right)\label{6,}
\end{align}
Fig.~\ref{Fig6} links colors with binary indices of (\ref{matrix}) for each of the above cases.

The subtlety is that the definition of $\otimes$ that occurs in ({\ref{14!})--(\ref{17!}) cannot define the same tensor product for all the matrices (\ref{4,})--(\ref{6,}). Rather, we obtain six different tensor products,
%11111111111111111111111111111111111111111111111111111111111111111111
\begin{align}
P_0\otimes_{123}\mathbb{I}
&=
P_C+P_M,
&P_0\otimes_{321}\mathbb{I}
&=
P_C+P_G,\label{22,.}\\
P_1\otimes_{123}\mathbb{I}
&=
P_Y+P_G,
&P_1\otimes_{321}\mathbb{I}
&=
P_M+P_Y,\\
\mathbb{I}\otimes_{123}P_0
&=
P_C+P_Y,
&\mathbb{I}\otimes_{321}P_0
&=
P_C+P_Y,\\
\mathbb{I}\otimes_{123}P_1
&=
P_M+P_G,
&\mathbb{I}\otimes_{321}P_1
&=
P_M+P_G,
\end{align}
%222222222222222222222222222222222222222222222222222222222222222222222
\begin{align}
P_0\otimes_{213}\mathbb{I}
&=
P_C+P_Y,
&P_0\otimes_{231}\mathbb{I}
&=
P_C+P_G,\\
P_1\otimes_{213}\mathbb{I}
&=
P_M+P_G,
&P_1\otimes_{231}\mathbb{I}
&=
P_M+P_Y,\\
\mathbb{I}\otimes_{213}P_0
&=
P_C+P_M,
&\mathbb{I}\otimes_{231}P_0
&=
P_C+P_M,\\
\mathbb{I}\otimes_{213}P_1
&=
P_Y+P_G,
&\mathbb{I}\otimes_{231}P_1
&=
P_Y+P_G,
\end{align}
%333333333333333333333333333333333333333333333333333333333333333333333333333
\begin{align}
P_0\otimes_{312}\mathbb{I}
&=
P_C+P_Y,
&P_0\otimes_{132}\mathbb{I}
&=
P_C+P_M,\\
P_1\otimes_{312}\mathbb{I}
&=
P_M+P_G,
&P_1\otimes_{132}\mathbb{I}
&=
P_Y+P_G,\\
\mathbb{I}\otimes_{312}P_0
&=
P_C+P_G,
&\mathbb{I}\otimes_{132}P_0
&=
P_C+P_G,\\
\mathbb{I}\otimes_{312}P_1
&=
P_M+P_Y,
&\mathbb{I}\otimes_{132}P_1
&=
P_M+P_Y.\label{33,.}
\end{align}
Notice that all these projectors commute.

The awkward indices in $\otimes_{abc}$ become clearer if we introduce unitary operators that  permute 
\be
|0\rangle &=& |00\rangle=|C\rangle,\label{7'}\\
|1\rangle &=&|01\rangle=|M\rangle,\\
|2\rangle &=&|10\rangle=|Y\rangle,\\
|3\rangle &=&|11\rangle=|G\rangle.\label{10'}
\ee
into
\be
|0\rangle &=& U_{abc}|00\rangle=|00_{abc}\rangle=|0\rangle\otimes_{abc}|0\rangle,\label{7,}\\
|a\rangle &=&U_{abc}|01\rangle=|01_{abc}\rangle=|0\rangle\otimes_{abc}|1\rangle,\\
|b\rangle &=&U_{abc}|10\rangle=|10_{abc}\rangle=|1\rangle\otimes_{abc}|0\rangle,\\
|c\rangle &=&U_{abc}|11\rangle=|11_{abc}\rangle=|1\rangle\otimes_{abc}|1\rangle.\label{10,}
\ee
Formulas (\ref{7,})--(\ref{10,}) are the definitions of $|rs_{abc}\rangle$ and $|r\rangle\otimes_{abc}|s\rangle$.
Formulas (\ref{7'})--(\ref{10'}) should not (yet) be understood in terms of a tensor structure --- for the moment this is just a binary parametrization of the first four whole numbers.
However, the consistency condition
\be
|0\rangle &=& U_{123}|00\rangle=|00_{123}\rangle=|0\rangle\otimes_{123}|0\rangle,\label{7,,}\\
|1\rangle &=&U_{123}|01\rangle=|01_{123}\rangle=|0\rangle\otimes_{123}|1\rangle,\\
|2\rangle &=&U_{123}|10\rangle=|10_{123}\rangle=|1\rangle\otimes_{123}|0\rangle,\\
|3\rangle &=&U_{123}|11\rangle=|11_{123}\rangle=|1\rangle\otimes_{123}|1\rangle,\label{10,,}
\ee
shows that $U_{123}$ is the identity operator and thus one can think of 
\be
|rs\rangle=|rs_{123}\rangle=|r\rangle\otimes_{123}|s\rangle\label{31''}
\ee
in terms of ``the'' ``ordinary'' tensor product. In fact, in matrix representations we will identify $\otimes_{123}=\otimes$ with the {\tt KroneckerProduct} matrix operation employed in Wolfram Mathematica.

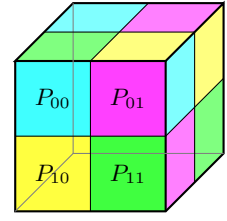
\begin{figure}
\begin{tikzpicture}
    % Define shadow shading
    \definecolor{mycyan}{rgb}{0.0, 1.0, 1.0}
    \definecolor{mymagenta}{rgb}{1.0, 0.0, 1.0}
    \definecolor{myyellow}{rgb}{1.0, 1.0, 0.0}
    \definecolor{mygreen}{rgb}{0.0, 1.0, 0.0}
    
    % Front Face Grid Cells
    \fill[mycyan, opacity=.75] (0,1,2) -- (1,1,2) -- (1,2,2) -- (0,2,2) -- cycle; % Top Left
    \fill[mymagenta, opacity=.75] (1,1,2) -- (2,1,2) -- (2,2,2) -- (1,2,2) -- cycle; % Top Right
    \fill[myyellow, opacity=.75] (0,0,2) -- (1,0,2) -- (1,1,2) -- (0,1,2) -- cycle; % Bottom Left
    \fill[mygreen, opacity=.75] (1,0,2) -- (2,0,2) -- (2,1,2) -- (1,1,2) -- cycle; % Bottom Right
    
    % Top Face Grid Cells
    \fill[mygreen, opacity=.5] (0,2,1) -- (1,2,1) -- (1,2,2) -- (0,2,2) -- cycle; % Top Left
    \fill[myyellow, opacity=.5] (1,2,1) -- (2,2,1) -- (2,2,2) -- (1,2,2) -- cycle; % Top Right
    \fill[mycyan, opacity=.5] (0,2,0) -- (1,2,0) -- (1,2,1) -- (0,2,1) -- cycle; % Bottom Left
    \fill[mymagenta, opacity=.5] (1,2,0) -- (2,2,0) -- (2,2,1) -- (1,2,1) -- cycle; % Bottom Right
    
    % Right Face Grid Cells
    \fill[mycyan, opacity=.5] (2,1,1) -- (2,1,2) -- (2,2,2) -- (2,2,1) -- cycle; % Top Left
    \fill[mymagenta, opacity=.5] (2,0,1) -- (2,0,2) -- (2,1,2) -- (2,1,1) -- cycle; % Bottom Right
    \fill[myyellow, opacity=.5] (2,1,0) -- (2,1,1) -- (2,2,1) -- (2,2,0) -- cycle; % Top Left
    \fill[mygreen, opacity=.5] (2,0,0) -- (2,0,1) -- (2,1,1) -- (2,1,0) -- cycle; % Bottom Right
    
    % Draw Cube
    \draw[thick] (2,2,0)--(0,2,0)--(0,2,2)--(2,2,2)--(2,2,0)--(2,0,0)--(2,0,2)--(0,0,2)--(0,2,2);
    \draw[thick] (2,2,2)--(2,0,2);
    \draw[gray]  (2,0,0)--(0,0,0)--(0,2,0);
    \draw[gray]  (0,0,0)--(0,0,2);
    
    % Draw Grids
    % Grid on front face
    \draw (1,0,2) -- (1,2,2);
    \draw (0,1,2) -- (2,1,2);
    % Grid on right face
    \draw (2,1,0) -- (2,1,2);
    \draw (2,0,1) -- (2,2,1);
    % Grid on top face
    \draw (1,2,0) -- (1,2,2);
    \draw (2,2,1) -- (0,2,1);
    
    % Labeling Cells on front face
    \draw (0.5,1.5,2) node {$P_{00}$};
    \draw (1.5,1.5,2) node {$P_{01}$};
    \draw (0.5,0.5,2) node {$P_{10}$};
    \draw (1.5,0.5,2) node {$P_{11}$};
\end{tikzpicture}

\caption{The unstable decaying system in a superposition of four orthogonal colour states CYGM can be also represented as an abstract cube. This decay leads to six inequivalent local products. Thus, one can associate six different tensor products $\otimes_{abc}$ with six faces of the abstract cube. Classically detection of results happens only on one chosen face of the cube. The global information about colours is lost during detection of results by Alice and Bob but leads to inequivalency of tensor products on the faces. One can transition among the faces applying rotations $U_{abc}$.
Interestingly, what is entangled on one face in a chosen tensor algebra can be separable on the another face with its tensor algebra. }
\label{Figcube}
\end{figure}

The matrix elements, $(U_{abc})_{kl}=\langle k|U_{abc}|l\rangle$, $k,l=0\dots 3$,  can be collected in six matrices,
\begin{align}
&(U_{123})_{kl}
=\left(
\begin{array}{cccc}
1 & 0 & 0 & 0\\
0 & 1 & 0 & 0\\
0 & 0 & 1 & 0\\
0 & 0 & 0 & 1
\end{array}
\right) 
&(U_{321})_{kl}
=\left(
\begin{array}{cccc}
1 & 0 & 0 & 0\\
0 & 0 & 0 & 1\\
0 & 0 & 1 & 0\\
0 & 1 & 0 & 0
\end{array}
\right) \label{27,}\\
&(U_{213})_{kl}
=\left(
\begin{array}{cccc}
1 & 0 & 0 & 0\\
0 & 0 & 1 & 0\\
0 & 1 & 0 & 0\\
0 & 0 & 0 & 1
\end{array}
\right) 
&(U_{231})_{kl}
=\left(
\begin{array}{cccc}
1 & 0 & 0 & 0\\
0 & 0 & 0 & 1\\
0 & 1 & 0 & 0\\
0 & 0 & 1 & 0
\end{array}
\right) \\
&(U_{312})_{kl}
=\left(
\begin{array}{cccc}
1 & 0 & 0 & 0\\
0 & 0 & 1 & 0\\
0 & 0 & 0 & 1\\
0 & 1 & 0 & 0
\end{array}
\right) 
&(U_{132})_{kl}
=\left(
\begin{array}{cccc}
1 & 0 & 0 & 0\\
0 & 1 & 0 & 0\\
0 & 0 & 0 & 1\\
0 & 0 & 1 & 0
\end{array}
\right)\label{29,}
\end{align}
Each matrix $(U_{abc})_{kl}$ is indexed by the permutation $(0,1,2,3)\to(0,a,b,c)$ it represents, and its position in the list (\ref{27,})--(\ref{29,}) matches an appropriate diagram from Fig.~\ref{Fig6}. We then find
\be
A\otimes_{abc}B &=& 
U_{abc}(A\otimes_{123}B)U_{abc}^\dag.\label{AabcB}
\ee
More generally, 
\be
A\otimes_{a''b''c''}B &=& 
U_{a'b'c'}(A\otimes_{abc}B)U_{a'b'c'}^\dag,\label{AabcB'}
\ee
where 
\be
U_{a''b''c''}=U_{a'b'c'}U_{abc},
\ee
since a composition of two permutations is a permutation. Moreover, if 
\be
(A\otimes_{abc}B)(C\otimes_{abc}D)=(AC)\otimes_{abc}(BD)\label{ABCD}
\ee
is true for some $\otimes_{abc}$, then it is true of any $\otimes_{abc}$. However, in general
\be
(A\otimes_{abc}B)(C\otimes_{a'b'c'}D)\neq (AC)\otimes_{a''b''c''}(BD)\label{ABCD'},
\ee
unless $(a,b,c)=(a',b',c')=(a'',b'',c'')$. 

Accordingly, there is nothing special in $\otimes_{123}$, so that any $\otimes_{abc}$ can be identified in matrix formulas with Wolfram Mathematica's {\tt KroneckerProduct}. We conclude that as any matrix is a table of coordinates of a  vector in some concrete basis, the identification of $\otimes_{123}=\otimes$ with {\tt KroneckerProduct} plays the same formal role. When it comes to concrete calculations with concrete operators, such as the Pauli-matrices representation of spin-1/2, we can use the usual matrix ``drag-and-drop'' Kronecker product $\otimes$. However, we have to be aware that at the level of tensor products, such an $\otimes$ can represent any $\otimes_{abc}$.

\subsection*{Correlations in space become correlata in time --- duality of quantum entanglement}

Before we proceed further, let us explain our terminology related to projectors, propositions, eigenvalues, truth values, and bits. The latter three notions have numerical values 0 and 1, which may lead to misunderstandings.

Any projector is self-adjoint and thus represents a quantum observable, in principle measurable in some experiment. From a quantum-logic perspective a projector represents a proposition (statement about the truth value of a given sentence). Projectors have eigenvalues 0 and 1, corresponding to the truth values of the corresponding propositions. For example, a measurement of an observable $P$ representing proposition ``Alice {\tt AND} Bob'' produces 0 if the proposition ``Alice {\tt AND} Bob'' is false. The corresponding eigenvector of $P$ is then a general superposition of $|00\rangle$, 
$|01\rangle$, and $|10\rangle$. Spectral representation of the projector is $P=|11\rangle\langle 11|$ because the truth value of the proposition ``1 {\tt AND} 1'' equals 1, and thus the eigenvalue of $P$ must be 1 as well, a fact happening only in case the corresponding eigenvector is $|11\rangle$.

The logical negation of proposition $P$ is given by $\mathbb{I}-P$, where $\mathbb{I}$ is the identity operator. In this paper, the pairs $P_0$ and $P_1$ are related by negation. Whenever we say that $P_0$ represents a zero bit, we mean that a measurement of $P_0$ returns 1 if the proposition ``bit equals 0'' is true. So, a measurement of $P_0$ returns 0 if the proposition ``bit equals 0'' is false. Analogously, $P_1$ ``represents a unit bit'' means that a measurement of $P_1$ returns 1 if the proposition ``bit equals 1'' is true. A measurement of $P_1$ returns 0 if the proposition ``bit equals 1'' is false. 

The above terminology is consistent with probability interpretation: $p_0=\langle\psi|P_0|\psi\rangle$ is the probability of finding 0 in a measurement performed in a physical system whose state is given by $|\psi\rangle$.

Now, let us return to the six matrices (\ref{4,})--(\ref{6,}). We will concentrate on three of them, because they appear in examples. Let us rewrite them as follows:
\be
\left(
\begin{array}{cc}
P_{00} & P_{01}\\
P_{10} & P_{11}
\end{array}
\right)
&=&
\left(
\begin{array}{cc}
P_{00_{123}} & P_{01_{123}}\\
P_{10_{123}} & P_{11_{123}}
\end{array}
\right),\\
\left(
\begin{array}{cc}
P_{00} & P_{11}\\
P_{10} & P_{01}
\end{array}
\right)
&=&
\left(
\begin{array}{cc}
P_{00_{321}} & P_{01_{321}}\\
P_{10_{321}} & P_{11_{321}}
\end{array}
\right)\label{31,}
\\
\left(
\begin{array}{cc}
P_{00} & P_{10}\\
P_{01} & P_{11}
\end{array}
\right)
&=&
\left(
\begin{array}{cc}
P_{00_{213}} & P_{01_{213}}\\
P_{10_{213}} & P_{11_{213}}
\end{array}
\right).
\ee
Local bits of subsystems Alice$_{abc}$ and Bob$_{abc}$ are defined by the usual rule
 \be
P_0\otimes_{abc}\mathbb{I}
&=& 
P_{00_{abc}}+ P_{01_{abc}},\label{37'}\\
P_1\otimes_{abc}\mathbb{I}
&=& 
P_{10_{abc}}+ P_{11_{abc}},\\
\mathbb{I}\otimes_{abc}P_0
&=&
P_{00_{abc}}+ P_{10_{abc}},\\
\mathbb{I}\otimes_{abc}P_1
&=&
P_{01_{abc}}+ P_{11_{abc}}.\label{40'}
\ee
The formulas become less natural if we rewrite (\ref{37'})--(\ref{40'}) by means of the parametrization
\be
\left(
\begin{array}{cc}
P_C & P_M\\
P_Y & P_G
\end{array}
\right)
=
\left(
\begin{array}{cc}
P_{00} & P_{01}\\
P_{10} & P_{11}
\end{array}
\right)\label{34,}
\ee
that refers to $\otimes=\otimes_{123}$.  The choice (\ref{34,}) is completely arbitrary, of course, and should be regarded as an arbitrary ``system of coordinates'' for all the other decompositions into subsystems. 
As we will see in a moment, referring all the projectors to our reference labeling (\ref{34,}) leads to a  ``space-time'' interpretation of the six tensor products.

There are no controversies over the definition of $\otimes_{123}$,
\begin{align}
P_0\otimes_{123}\mathbb{I}
&=
P_{00}+P_{01}, \textrm{(0 of Alice)}
\\
P_1\otimes_{123}\mathbb{I}
&=
P_{10}+P_{11},\textrm{(1 of Alice)}
\\
\mathbb{I}\otimes_{123}P_0
&=
P_{00}+P_{10},\textrm{(0 of Bob)}
\\
\mathbb{I}\otimes_{123}P_1
&=
P_{01}+P_{11},\textrm{(1 of Bob)}
\end{align}
We will sometimes refer to Alice and Bob defined by the above formulas as Alice$_{123}$ and Bob$_{123}$.
Tensor product $\otimes_{123}$ is here identified with the Kronecker product of matrices and is equivalent to typical $\otimes$ being our reference system. Hence, e.g.
\be
P_0\otimes_{123}\mathbb{I}
&=&
\left(
\begin{array}{cc}
1 & 0\\
0 & 0
\end{array}
\right)
\otimes \mathbb{I}
=
\left(
\begin{array}{cccc}
1 & 0 & 0 & 0\\
0 & 1 & 0 & 0\\
0 & 0 & 0 & 0\\
0 & 0 & 0 & 0
\end{array}
\right),
\ee
An interpretation of the right bit in the tensor product $\otimes_{321}$ is unchanged with respect to $\otimes_{123}$, but the left bit carries information about the correlation between Alice$_{123}$, and Bob$_{123}$, 
\begin{align}
P_0\otimes_{321}\mathbb{I}
&=
P_{00}+P_{11},\textrm{(Alice {\tt IFF} Bob)}\label{AB0011}\\
P_1\otimes_{321}\mathbb{I}
&=
P_{01}+P_{10},\textrm{(Alice {\tt XOR} Bob)}\\
\mathbb{I}\otimes_{321}P_0
&=
P_{00}+P_{10},\textrm{(0 of Bob)}\\
\mathbb{I}\otimes_{321}P_1
&=
P_{01}+P_{11},\textrm{(1 of Bob)}
\end{align}
Although the ``right'' bit correctly illustrates the position of the bit in $P_{\alpha\beta}$, the term ``left'' may be somewhat misleading: Projector $P_0\otimes_{321}\mathbb{I}$ represents the logical equivalence {\tt IFF} (if and only if),
\be
\textrm{Alice}_{321}=(\textrm{Alice$_{123}$ {\tt IFF} Bob$_{123}$})
\ee 

$\otimes_{321}$ carries hidden global information about source of the local states on Alice' and Bob's sites. Equation (\ref{AB0011}) carries information that local 0 bit on Alice' site has a source in cyan and green coloured states.

At the level of matrix representations the rule that links bits defined by $\otimes_{321}$ and $\otimes$ is,
\be
P_0\otimes_{321}\mathbb{I}
&=&
U_{321}(P_0\otimes\mathbb{I})U_{321}^\dag, \quad\textrm{etc.}
\ee
The unitary operator $U_{321}$ interchanges $01$ and $11$,
\be
U_{321}
=
|00\rangle\langle 00|
+
|01\rangle\langle 11|
+
|10\rangle\langle 10|
+
|11\rangle\langle 01|.
\ee
Explicitly,
\be
U_{321}P_{00}U_{321}^\dag
=
P_{00_{321}}
=
U_{321}|00\rangle\langle 00|U_{321}^\dag=P_{00},&\\
U_{321}P_{01}U_{321}^\dag
=
P_{01_{321}}
=
U_{321}|01\rangle\langle 01|U_{321}^\dag=P_{11},&\\
U_{321}P_{10}U_{321}^\dag
=
P_{10_{321}}
=
U_{321}|10\rangle\langle 10|U_{321}^\dag=P_{10},&\\
U_{321}P_{11}U_{321}^\dag
=
P_{11_{321}}
=
U_{321}|11\rangle\langle 11|U_{321}^\dag=P_{01}.&
\ee
$U_{321}$ is the {\tt CNOT} operation performed on the left bit, and controlled by the right bit. On the other hand, 
 the role of $U_{abc}$ is to relabel the basis. 

\begin{figure}
\includegraphics[width=6 cm]{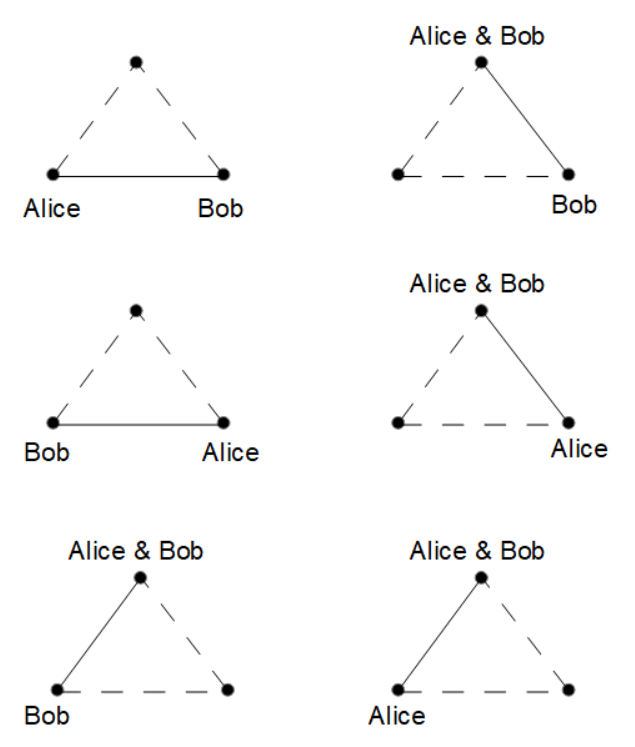}
\caption{Six types of correlations described by six types of bits, associated with six decompositions into subsystems, and six tensor products. Effectively, there are three types of questions associated with alternative tensor products: ``Are the bits of Alice and Bob different?'', ``What is the bit of Alice?", and ``What is the bit of Bob?". The three types occur in two versions, with Alice and Bob interchanged. Joint questions about Alice and Bob are causally related with separate questions about Alice and Bob. The four tensor structures are thus related to causal correlations (``correlations in time''). The joint questions can refer either to the past (correlations at the source), or to the future (joint detections). In Bell-type experiments \cite{Bell} both cases occur simultaneously. The arrow of time in this figure can point from the bottom to the top, or from the top to the bottom. In Mermin's terminology, a correlation in $\otimes_{123}$ and $\otimes_{213}$, becomes a correlatum in $\otimes_{321}$, $\otimes_{231}$, $\otimes_{312}$, and $\otimes_{132}$, and vice versa.}
\label{Fig_c}
\end{figure}

As the ``right'' bit of Bob does not change its interpretation, one might have the impression that local operators associated with Bob should remain unchanged, which is not the case as it turns out.
Let us discuss the problem in more detail. Let us recall that $|\alpha\beta_{321}\rangle=U_{321}|\alpha\beta\rangle=U_{321}|\alpha\beta_{123}\rangle$, so
%\begin{widetext}
\be
\mathbb{I}\otimes_{321}\sigma_3
&=&
U_{321}\big(\mathbb{I}\otimes(|0\rangle\langle 0|-|1\rangle\langle 1|)\big)U_{321}^\dag
\nonumber\\
&=&
|00\rangle\langle 00|-|11\rangle\langle 11|+|10\rangle\langle 10|-|01\rangle\langle 01|
\nonumber\\
&=&
\mathbb{I}\otimes\sigma_3,
\ee
which indeed shows that eigenvectors of $\sigma_3$ retain their interpretation independently of the choice of $\otimes_{321}$ or $\otimes_{123}=\otimes$. However, the situation changes when we consider the remaining Pauli matrices. For example the local {\tt NOT} operation in the subsystem  Bob$_{321}$ reads
\be
\mathbb{I}\otimes_{321}\sigma_1
&=&
U_{321}\big(\mathbb{I}\otimes(|0\rangle\langle 1|+|1\rangle\langle 0|)\big)U_{321}^\dag
\nonumber\\
&=&
\big(|0\rangle\langle 1|+|1\rangle\langle 0|\big)\otimes \big(|0\rangle\langle 1|+|1\rangle\langle 0|\big)
\nonumber\\
&=&
\sigma_1\otimes\sigma_1=\sigma_1\otimes_{123}\sigma_1.
\ee
Now, let us take a look at the negation of the left bit,
\be
\sigma_1\otimes_{321}\mathbb{I}
&=&
U_{321}\big((|0\rangle\langle 1|+|1\rangle\langle 0|)\otimes\mathbb{I}\big)U_{321}^\dag
\\
&=&
\big(|0\rangle\langle 1|+|1\rangle\langle 0|\big)\otimes \big(|0\rangle\langle 0|+ |1\rangle\langle 1|\big)
\\
&=&
\sigma_1\otimes\mathbb{I}.
\ee
One analogously checks that
\be
\sigma_3\otimes_{321}\mathbb{I}
&=&
\sigma_3\otimes\sigma_3.
\ee
The four formulas,
\be
\mathbb{I}\otimes_{321}\sigma_3
&=&
\mathbb{I}\otimes\sigma_3,\\
\mathbb{I}\otimes_{321}\sigma_1
&=&
\sigma_1\otimes\sigma_1,\\
\sigma_1\otimes_{321}\mathbb{I}
&=&
\sigma_1\otimes\mathbb{I},\\
\sigma_3\otimes_{321}\mathbb{I}
&=&
\sigma_3\otimes\sigma_3,\label{91!!!}
\ee
seem inconsistent at first glance, but they illustrate the change of $\otimes_{123}$-correlations into  $\otimes_{321}$-correlata.
Indeed, eigenvectors of
\be
P_0\otimes_{321}\mathbb{I}&=&
\frac{1}{2}(\mathbb{I}+\sigma_3)\otimes_{321}\mathbb{I}
\\
&=&
\frac{1}{2}\mathbb{I}\otimes\mathbb{I}+\frac{1}{2}\sigma_3\otimes\sigma_3
\\
&=&
\left(
\begin{array}{cccc}
1 & 0 & 0 & 0\\
0 & 0 & 0 & 0\\
0 & 0 & 0 & 0\\
0 & 0 & 0 & 1
\end{array}
\right)=P_{00}+P_{11}
\ee
are given by
\begin{align}
|f_\parallel\rangle
&=
f_{00}|00\rangle+f_{11}|11\rangle &\textrm{($\otimes$-entangled state)}\label{ent92}\\
&=
f_{00}|00_{321}\rangle+f_{11}|01_{321}\rangle
&\textrm{($\otimes_{321}$-product state)}\label{ent93}
\\
|f_\perp\rangle
&=
f_{01}|01\rangle+f_{10}|10\rangle &\textrm{($\otimes$-entangled state)}\\
&=
f_{01}|11_{321}\rangle+f_{10}|10_{321}\rangle &\textrm{($\otimes_{321}$-product state)}
\end{align}
with
\be
(P_0\otimes_{321}\mathbb{I})|f_\parallel\rangle
&=&|f_\parallel\rangle,\\
(P_1\otimes_{321}\mathbb{I})|f_\parallel\rangle
&=&0,\\
(P_0-P_1)\otimes_{321}\mathbb{I})|f_\parallel\rangle
&=&
|f_\parallel\rangle
=
(\sigma_3\otimes_{321}\mathbb{I})|f_\parallel\rangle,\\
(P_0\otimes_{321}\mathbb{I})|f_\perp\rangle
&=&0,\\
(P_1\otimes_{321}\mathbb{I})|f_\perp\rangle
&=&|f_\perp\rangle,\\
(P_0-P_1)\otimes_{321}\mathbb{I})|f_\perp\rangle
&=&
-|f_\perp\rangle
=(\sigma_3\otimes_{321}\mathbb{I})|f_\perp\rangle,
\ee
which shows that
\be
(\sigma_3\otimes_{321}\mathbb{I})|f_\parallel\rangle &=& |f_\parallel\rangle,\label{100'}\\
(\sigma_3\otimes_{321}\mathbb{I})|f_\perp\rangle &=& -|f_\perp\rangle.\label{101'}
\ee
The particular cases of (\ref{100'})--(\ref{101'}) are
\begin{widetext}
\be
(\sigma_3\otimes_{321}\mathbb{I})|00_{321}\rangle &=& |00_{321}\rangle=|00\rangle=(\sigma_3\otimes_{321}\mathbb{I})|00\rangle=(\sigma_3\otimes\sigma_3)|00\rangle,\\
(\sigma_3\otimes_{321}\mathbb{I})|01_{321}\rangle &=& |01_{321}\rangle=|11\rangle=(\sigma_3\otimes_{321}\mathbb{I})|11\rangle=(\sigma_3\otimes\sigma_3)|11\rangle,\\
(\sigma_3\otimes_{321}\mathbb{I})|11_{321}\rangle &=& -|11_{321}\rangle=-|01\rangle=(\sigma_3\otimes_{321}\mathbb{I})|01\rangle=(\sigma_3\otimes\sigma_3)|01\rangle,\\
(\sigma_3\otimes_{321}\mathbb{I})|10_{321}\rangle &=& -|10_{321}\rangle=-|10\rangle=(\sigma_3\otimes_{321}\mathbb{I})|10\rangle=(\sigma_3\otimes\sigma_3)|10\rangle,
\ee
which explicitly cross-checks (\ref{91!!!}).
Analogously, one verifies that the two forms of negation of the left bit are indeed equivalent:
\be
(\sigma_1\otimes_{321}\mathbb{I})(f_{00}|00_{321}\rangle+f_{11}|01_{321}\rangle)
&=&
f_{00}|10_{321}\rangle+f_{11}|11_{321}\rangle,\\
(\sigma_1\otimes\mathbb{I})
(f_{00}|00\rangle+f_{11}|11\rangle)
&=&
f_{00}|10\rangle+f_{11}|01\rangle\\
&=&
f_{00}|10_{321}\rangle+f_{11}|11_{321}\rangle.
\ee
\end{widetext}
Equations (\ref{ent92}) and (\ref{ent93}) show something peculiar as if an entangled state was at the same time a product state. This apparent paradox originates from the relativity of tensor product structures. The equation (\ref{ent93}) stores hidden global information about the source of the correlations and stores information about temporal correlations (and causal relation between Alice and Bob) which is not manifested at the level of purely spatial correlations between Alice and Bob for the plain $\otimes$. This phenomenon exemplifies the \textit{duality of quantum entanglement} whose structure also varies based on the chosen tensor algebra associated with observers. 

Projectors such as $P_{\alpha}\otimes_{321}\mathbb{I}$ represent observables that can be measured only in case we know simultaneously the results of local measurements performed in subsystems Alice$_{123}$ and Bob$_{123}$, a typical situation in Bell-type experiments or tests for eavesdropping in entangled-state quantum cryptography. Notice that states from the Bell basis,
\be
|\Psi_\pm\rangle &=&
\frac{1}{\sqrt{2}}(|01\rangle\pm |10\rangle),\\
|\Phi_\pm\rangle &=&
\frac{1}{\sqrt{2}}(|00\rangle\pm |11\rangle)
\ee
satisfy
\be
(P_0\otimes_{321}\mathbb{I})|\Psi_\pm\rangle &=& 0,\\
(P_1\otimes_{321}\mathbb{I})|\Psi_\pm\rangle &=& |\Psi_\pm\rangle,\\
(P_0\otimes_{321}\mathbb{I})|\Phi_\pm\rangle &=& |\Phi_\pm\rangle,\\
(P_1\otimes_{321}\mathbb{I})|\Phi_\pm\rangle &=& 0.
\ee
Hence, a definite truth value of the proposition $P_{\alpha}\otimes_{321}\mathbb{I}$ can be determined either at the source of the pairs of particles (preselection: those who arrange the experiment know which state is prepared), hence in the past of the detection events,  or at the final stage when the results of local measurements performed by Alice$_{123}$ and Bob$_{123}$ are collected and thus can be compared with one another --- hence in the future of the detection events (postselection). In this sense $\otimes_{123}$  \textit{describes space-like correlations}, whereas $\otimes_{321}$ \textit{describes the time-like ones}.

Considering the singlet state $|\Psi_-\rangle$, we know that it is an entangled state with respect to the ``usual'' tensor product $\otimes_{123}$. However, with respect to $\otimes_{321}$ it must be a product state because  the truth value of the proposition (Alice$_{123}$ {\tt XOR} Bob$_{123}$) is 1, while at the same time the truth value of ``bit of Bob$_{321}$ is 0'' is either 0 or 1. And indeed,
\be
|01\rangle-|10\rangle
&=&
|0\rangle\otimes_{123}|1\rangle-|1\rangle\otimes_{123}|0\rangle\\
&=&
|11_{321}\rangle-|10_{321}\rangle\\
&=&
|1\rangle\otimes_{321}|1\rangle-|1\rangle\otimes_{321}|0\rangle\\
&=&
|1\rangle\otimes_{321}\big(|1\rangle-|0\rangle\big),\label{123!}
\ee
is entangled with respect to $\otimes_{123}$ and product with respect to $\otimes_{321}$. The singlet state is space-like entangled and time-like untangled.

\subsection*{Local change of basis}

Assume $|\alpha\beta_{abc}\rangle$ is an orthonormal basis. A local change of basis is defined by
\be
|\alpha_V\beta_{Wabc}\rangle
&=&
\sum_{\gamma,\delta}V_{\alpha\gamma}W_{\beta\delta}|\gamma\delta_{abc}\rangle\\
&=&
(V\otimes_{abc}W)|\alpha\beta_{abc}\rangle,
\ee
where the matrices $V_{\alpha\gamma}$ and $W_{\beta\delta}$ are unitary.

In order to formulate the Einstein-Podolsky-Rosen-Bohm argument \cite{EPR,Bohm} in our generalized framework, we have to discuss consequences of a change of basis in 
(\ref{1?})--(\ref{8?}) into
\be
|0_V0_{W123}\rangle &=& |0_V\rangle \otimes_{123} |0_W\rangle ,\label{1??}\\
|0_V1_{W123}\rangle &=& |0_V\rangle \otimes_{123} |1_W\rangle ,\\
|1_V0_{W123}\rangle &=& |1_V\rangle \otimes_{123} |0_W\rangle ,\\
|1_V1_{W123}\rangle &=& |1_V\rangle \otimes_{123} |1_W\rangle ,
\ee
and
\be
P_{0_V0_{W123}} &=& P_{0_V}\otimes_{123} P_{0_W},\\
P_{0_V1_{W123}} &=& P_{0_V}\otimes_{123} P_{1_W},\\
P_{1_V0_{W123}} &=& P_{1_V}\otimes_{123} P_{0_W},\\
P_{1_V1_{W123}} &=& P_{1_V}\otimes_{123} P_{1_W}.\label{8??}
\ee
The formula
\be
P_{\alpha_V\beta_{W123}}
&=&
(V\otimes_{123}W) P_{\alpha\beta_{123}}(V\otimes_{abc}W)^\dag,
\ee
trivially generalizes to any $abc$, 
\be
P_{\alpha_V\beta_{Wabc}}
&=&
(V\otimes_{123}W) P_{\alpha\beta_{abc}}(V\otimes_{123}W)^\dag,\label{163}
\ee
because $P_{\alpha\beta_{abs}}$ is related to $P_{\gamma\delta_{123}}$ by a permutation $\pi$ of  pairs of the Greek indices, $\alpha\beta\mapsto \gamma\delta=\pi(\alpha\beta)$, hence
\be
(\ref{163})
&=&
(V\otimes_{123}W) P_{\pi(\alpha\beta)_{123}}(V\otimes_{123}W)^\dag
\\
&=&
P_{\pi(\alpha_V\beta_W)_{123}}
=
P_{\alpha_V\beta_{Wabc}}.
\ee
Let us further note that (\ref{163}) gives an example of a map
\be
P_{\alpha}\otimes_{abc}P_{\beta}
\mapsto
U \big(P_{\alpha}\otimes_{abc}P_\beta \big)U^\dag
=
P_{\alpha_U}\otimes_{abc}P_{\beta_U}\nonumber,\\
\label{163!!}
\ee
that should {\it not\/} be regarded as a change $\otimes_{abc}\mapsto \otimes_{U abc}$,
\be
P_{\alpha}\otimes_{abc}P_{\beta}
\mapsto
P_{\alpha}\otimes_{U abc}P_\beta ,\label{163!!!}
\ee
of the tensor product itself.
So, in case of $U=V\otimes W$ we obtain
\be
P_{\alpha_U}\otimes P_{\beta_U}
\ee
and not
\be
P_{\alpha}\otimes_{U abc}P_\beta,
\ee

Subtleties of this sort are essential for the EPR argument, as seen from the perspective of generalized tensor products. The physical meaning of the result is intuitively obvious: in case of $U=V\otimes_{123} W$ one does not influence the splitting of the system into subsystems (Alice and Bob are intact), but only  changes bases in the subsystems. This is exactly what we need in order to formulate the EPR argument. Interestingly, what was not obvious from the very outset, {\it although the product structure refers to one of the six tensor products only, it automatically applies to all the remaining tensor products as well\/}.

If we additionally assume $V=W\in\mathrm{SU}(2)$, 
 the singlet state remains unchanged. Indeed, on the one hand,
\be
|01_{123}\rangle-|10_{123}\rangle
&=&
|11_{321}\rangle-|10_{321}\rangle\label{166}
\\
&=&
|1\rangle\otimes_{321}\big(|1\rangle-|0\rangle\big),\label{167}
\ee
while on the other,
\be
|01_{123}\rangle-|10_{123}\rangle
&=&
|0_V1_{V123}\rangle-|1_V0_{V123}\rangle\label{168}\\
&=&
|1_V1_{V321}\rangle-|1_V0_{V321}\rangle\label{169}\\
&=&
|1_V\rangle\otimes_{321}\big(|1_V\rangle-|0_V\rangle\big)\label{170}
\ee
At first glance, the  equality of (\ref{167}) and (\ref{170}) is surprising for two reasons. First of all, it shows that product states can be invariant under changes of basis, a property typically attributed to entangled states. Secondly, it shows that the correlatum at the sides of Alice$_{321}$ of (\ref{167})--(\ref{170}) (i.e. the correlation with respect to $\otimes_{123}$) changes with the change of basis. So, there are infinitely many correlations in the singlet state.

After a little thought, one realizes what happens. There are indeed infinitely many correlations --- projections of spin on the same direction $\vec a$ are in a one-to-one relation for any 
$\vec a$ if the two-particle state is the singlet. The whole analysis we have performed so far referred to the eigen-basis of $\sigma_3$ (or, more precisely, we took an arbitrary basis and then defined $\sigma_3=|0\rangle\langle 0|-|1\rangle\langle 1|$). This is why the left bit corresponding to the subsystem defined by means of $\otimes_{321}$ (i.e. the correlation for $\otimes_{123}$) might seem unique. Our analysis shows that the correlation with respect to $\otimes_{123}$ behaves as a typical qubit, if we regard it as a correlatum for $\otimes_{321}$. 

For a nontrivial $V$, equality (\ref{168})  is possible only in the singlet case. Accordingly, the formula
\be
|1\rangle\otimes_{321}\big(|1\rangle-|0\rangle\big)
&=&
|1_V\rangle\otimes_{321}\big(|1_V\rangle-|0_V\rangle\big),\label{175}
\ee
provides another unique characterization of the singlet state. Moreover, if (\ref{175}) is true for a single nontrivial $V$, it remains true for any $V\in\mathrm{SU}(2)$. 
Accordingly, for any unitary $V$, the singlet state is the eigenstate of
\be
V\sigma_3 V^\dag\otimes_{321}V\sigma_1 V^\dag,
\ee
corresponding to the eigenvalue $+1$, a fact that generalizes (\ref{147?}).

\subsection*{Implications for the EPR argument}

Einstein-Podolsky-Rosen-Bohm argumentation is implicit in the following collection of equivalent formulas,
\be
|\Psi_-\rangle
&=&
\frac{1}{\sqrt{2}}
\big(|-_3\rangle\otimes_{123}|+_3\rangle - |+_3\rangle\otimes_{123}|-_3\rangle\big)\label{177!!}\\
&=&
\frac{1}{\sqrt{2}}
\big(|-_{3V}\rangle\otimes_{123}|+_{3V}\rangle - |+_{3V}\rangle\otimes_{123}|-_{3V}\rangle\big)\nonumber\\
\label{178!!}\\
&=&
\frac{1}{\sqrt{2}}|-_3\rangle\otimes_{321}\big(|-_3\rangle - |+_3\rangle\big)\label{179!!}\\
&=&
\frac{1}{\sqrt{2}}|-_{3V}\rangle\otimes_{321}\big(|-_{3V}\rangle - |+_{3V}\rangle\big)\label{180!!}\\
&=&
\frac{1}{\sqrt{2}}\big(|-_3\rangle - |+_3\rangle\big)\otimes_{231}|-_{3V}\rangle\label{181!!}\\
&=&
\frac{1}{\sqrt{2}}\big(|-_{3V}\rangle - |+_{3V}\rangle\big)\otimes_{231}|-_{3V}\rangle.\label{182!!}
\ee
Here,
\be
\sigma_3 &=& |+_3\rangle\langle +_3|-|-_3\rangle\langle -_3|,\\
\sigma_1 &=& |+_3\rangle\langle -_3|+|-_3\rangle\langle +_3|,\\
\sigma_{3V} &=& V\sigma_3 V^\dag,\\
\sigma_{1V} &=& V\sigma_1 V^\dag,
\ee
and $V\in \mathrm{SU}(2)$ is arbitrary.

The argumentation based on (\ref{177!!})--(\ref{178!!}) is well known. Formulas  (\ref{179!!})--(\ref{180!!}) and (\ref{181!!})--(\ref{182!!}) involve either Alice$_{231}$ or Bob$_{321}$, but never both Alice and Bob at the same time. The other qubit is the correlation between Alice$_{123}$ and  Bob$_{123}$. Notice that in the latter two cases the qubits of Alice$_{231}$ or Bob$_{321}$ are always described by {\it pure} states, because the singlet is a product state with respect to $\otimes_{321}$ and $\otimes_{231}$.

The EPR argument, when reformulated on the basis of (\ref{179!!})--(\ref{180!!}) is somewhat surprising, but reflects a tacit element of the standard argumentation. For let us begin with (\ref{179!!}). Bob$_{321}$ performs his measurement of $\sigma_3$ and finds $+1$ or $-1$. The fact that the result of Alice$_{321}$ is always $-1$ means that the result of a measurement performed by Alice$_{123}$ (not Alice$_{321}$!) will be exactly opposite to the one found by Bob$_{321}$. So, knowing the result obtained by Bob$_{321}$, and taking into account that the result obtained by Alice$_{321}$ is always the same (namely $-1$; hence the product structure of $|\Psi_-\rangle$) we automatically know what will be the result of the measurement of $\sigma_3$ if performed by Alice$_{123}$. This is why the vertices of the triangles in Fig.~\ref{Fig_c} correspond to simultaneously measurable correlata and correlations (recall  that the right-hand-sides of (\ref{22,.})--(\ref{33,.}) commute with one another).

If we repeat the same reasoning with (\ref{180!!}), the only difference is that $\sigma_3$ is now replaced by $\sigma_{3V}$. The result obtained by Alice$_{321}$ is again always the same, namely $-1$. But now it means that the result found by Alice$_{123}$ will be exactly opposite to the one found by Bob$_{321}$, if they both decide to measure $\sigma_{3V}$.

\section*{Conclusions}

The duality of correlations and correlata seems to be an overlooked aspect of the well known relativity of tensor structures. Quantum relativity of time and space appears here in a completely different light, suggesting that correlations may constitute the fabric of space-time itself.

The discussion is restricted to two qubits which allowed us to separate the tensor ambiguities from those that occur in Hilbert spaces of other non-prime dimensions \cite{Zanardi,Lloyd}. Moreover, the fact that we restrict the ambiguity to permutations of the four basis vectors, restricts the automorphism freedom in the sense of \cite{Kus,Thirring} to the permutation group only, making it a kind of fundamental gauge symmetry of any quantum theory, and simultaneously guaranteeing that the alternative tensor structures are mutually coexisting ($P_\alpha\otimes_{abc}\mathbb{I}$ commute with one another for any $a,b,c$). One should mention here that there exists yet another form of tensor ambiguity, related to modular tensor products \cite{MC2003}, which is not directly related to the above problems and applies to Hilbert spaces of any dimension.

The old problem of the borderland between an observer and that which is observed should be, perhaps, discussed anew from the perspective of alternative tensor structures. It is in principle possible that correlations within a system should be formulated by means of, say, $\otimes_{123}$, whereas those between the system and an observer should involve a tensor product analogous to $\otimes_{321}$ which turns entangled states into product ones, thus allowing ``system--universe'' product states $|\psi_s\rangle\otimes_{321}|\psi_s\rangle$ maintain system--observer correlations. 

Maybe this is the formal reason why predictions of Schr\"odinger equations can lead to observable consequences?

\section*{Acknowledgments}

We are indebted to Maciej Cichosz, Marek Kuś, Kamil Nalikowski, Marcin Pawłowski, Michał Szczepanik, and Karol Życzkowski for their remarks at various stages of this project. Some calculations were carried out at the Academic Computer Center in Gdańsk.

\section*{Appendix}\label{Appendix}

\subsection{Subsystems defined by $\otimes_{213}$}

Tensor product $\otimes_{213}$ leads to bits defined by
\begin{align}
P_0\otimes_{213}\mathbb{I}
&=
P_{00}+P_{10},\textrm{(0 of Bob)}
,\\
P_1\otimes_{213}\mathbb{I}
&=
P_{01}+P_{11},\textrm{(1 of Bob)}
\\
\mathbb{I}\otimes_{213}P_0
&=
P_{00}+P_{01},\textrm{(0 of Alice)}
\\
\mathbb{I}\otimes_{213}P_1
&=
P_{10}+P_{11}.\textrm{(1 of Alice)}
\end{align}
Hence, Alice$_{213}$$~=~$Bob$_{123}$, Bob$_{213}$~=~Alice$_{123}$, i.e.
\be
P_{\alpha}\otimes_{213}\mathbb{I} &=& \mathbb{I}\otimes_{123}P_{\alpha},\\
\mathbb{I}\otimes_{213}P_{\alpha} &=& P_{\alpha}\otimes_{123}\mathbb{I},\\
\sigma_\mu\otimes_{213}\sigma_\nu &=&\sigma_\nu\otimes_{123}\sigma_\mu.
\ee
One may wonder if the resulting decomposition into subsystems should be treated as different from the one implied by $\otimes_{123}$. If not, we should identify systems that differ only by permutations of Alice and Bob, which would further reduce the number of different tensor products to three.

\subsection{Subsystems defined by $\otimes_{231}$}

Tensor product $\otimes_{231}$ is similar to $\otimes_{321}$, but now Bob$_{231}$~=~Alice$_{123}$~=~Bob$_{213}$,
\begin{align}
P_0\otimes_{231}\mathbb{I}
&=
P_{00}+P_{11},\textrm{(Alice {\tt IFF} Bob)}\\
P_1\otimes_{231}\mathbb{I}
&=
P_{01}+P_{10},\textrm{(Alice {\tt XOR} Bob)}\\
\mathbb{I}\otimes_{231}P_0
&=
P_{00}+P_{01},\textrm{(0 of Alice)}\\
\mathbb{I}\otimes_{231}P_1
&=
P_{10}+P_{11},\textrm{(1 of Alice)}
\end{align}
and
Alice$_{231}$~=~Alice$_{321}$.

\subsection{Singlet state once again}

The singlet state satisfies simultaneously
\be
(\sigma_3\otimes_{321}\mathbb{I})|\Psi_-\rangle
&=&
-|\Psi_-\rangle,\label{144!!!}\\
(\mathbb{I}\otimes_{321}\sigma_1)|\Psi_-\rangle
&=&
-|\Psi_-\rangle,\label{145!!!}\\
(\mathbb{I}\otimes_{231}\sigma_3)|\Psi_-\rangle
&=&
-|\Psi_-\rangle,\label{146!!!}\\
(\sigma_1\otimes_{231}\mathbb{I})|\Psi_-\rangle
&=&
-|\Psi_-\rangle,\label{147!!!}
\ee
and
\be
(\sigma_3\otimes_{321}\sigma_1)|\Psi_-\rangle
&=&
|\Psi_-\rangle,\label{147?}\\
(\sigma_1\otimes_{231}\sigma_3)|\Psi_-\rangle
&=&
|\Psi_-\rangle,\label{106,}
\ee
implying (here $\sigma_k|\pm_k\rangle=\pm|\pm_k\rangle$, $k=1,3$)
\be
|\Psi_-\rangle
&=&
|-_3\rangle \otimes_{321} |-_1\rangle\label{107,}\\
&=&
|-_1\rangle \otimes_{231} |-_3\rangle,\label{108,}
\ee
which agrees with (\ref{123!}). 

Bearing in mind that single qubits are defined as $|0\rangle=|+_3\rangle$, $|1\rangle=|-_3\rangle$,  we observe that (\ref{107,}) agrees with the structure of the observable $\sigma_3\otimes_{321}\sigma_1$ if we interpret 
the left qubit satisfying $\sigma_3|-_3\rangle=-|-_3\rangle$ as the positive answer to the question ``Are the bits of Alice and Bob different?"

\end{document}